\documentclass[aps,prd,showpacs,superscriptaddress,twocolumn,nofootinbib]{revtex4-1}
\usepackage{graphicx}
\usepackage{amsmath}
\usepackage{amssymb}
\usepackage{ulem}
\usepackage{bm}
\usepackage{float}
\usepackage[colorlinks,
linkcolor=blue,
anchorcolor=blue,
urlcolor=blue,
citecolor=blue,
]{hyperref}

\begin{document}
\title{Fluctuations of topological charge and chiral density in the early stage of high energy nuclear collisions}
\author{M. R. Jia}
\author{J. H. Liu}
\author{H. F. Zhang \footnote{Corresponding author. Email address: zhanghongfei@lzu.edu.cn}}
\author{M. Ruggieri \footnote{Corresponding author. Email address: ruggieri@lzu.edu.cn}}
\affiliation{School of Nuclear Science and Technology, Lanzhou University. 222 South Tianshui Road, Lanzhou 730000, China}

\date{\today}

\begin{abstract}
We study systematically the topological charge density and the chiral density correlations in 
the early stage of high energy nuclear collisions: the initial condition is given by the McLerran-Venugopalan  model and 
the evolution of the gluon fields is studied via the Classical Yang-Mills equations up to proper time $\tau\approx 1$ fm/c
for an SU(2) evolving Glasma.
Topological charge is related to the gauge invariant $\bm E \cdot \bm B$ where $\bm E$ and $\bm B$ denote the
color-electric and color-magnetic fields, while the chiral density is produced via the chiral anomaly of
Quantum Chromodynamics.
We study how the correlation lengths are related to the collision energy, 
and how the correlated domains grow up with proper time in the transverse plane for a boost invariant longitudinal expansion.
We estimate the correlation lengths of both quantities, that after a short transient results of the order of the typical energy scale of the model, namely the inverse of the
saturation scale. We estimate the proper time for the formation of a steady state in which the production of the
chiral density in the transverse plane per unit rapidity slows down, as well as the amount of chiral density that would be present at the switch time
between the Classical Yang-Mills evolution and the relativistic transport or hydro for the quark-gluon plasma phase. 
\end{abstract}
\pacs{12.38.Aw, 25.75.-q}
\keywords{}
\maketitle

\section{Introduction}

The study of the initial condition in high energy collisions is a difficult but interesting problem related to the physics of relativistic heavy ion collisions (RHICs), as well as to that of high energy proton-proton (pp) and proton-nucleus (pA) collisions. The dynamics of the central rapidity region is determined by the small Bjorken $x$ gluons before the collision where saturation 
takes place \cite{Mueller:1999wm,PhysRevD.49.2233,PhysRevD.49.3352,PhysRevD.50.2225}. 
In the saturation region the gluon occupation number is large enough that classical effective field theories (EFTs) can be 
used  \cite{PhysRevD.55.5414,PhysRevD.54.5463,PhysRevD.59.014015,Iancu_20011,Iancu_20012}: the 
Lorentz-contracted colliding nuclei are
idealized to fly along the light cone, with the large-$x$ partons behaving as static sources 
of the small-$x$ modes that make the color-glass condensate (CGC) fields inside the two nuclei, 
see Refs.\cite{PhysRevD.49.2233,PhysRevD.49.3352,PhysRevD.50.2225,doi:10.1146/annurev.nucl.010909.083629,mclerran2008brief,mclerran2008color,GELISCOLOR} for reviews. 
Right after the collision, color sources form on the two collision sheets as a result of the interaction of the two
CGC colliding sheets,
in such a way longitudinal color electric, $\bm E$, and color magnetic, $\bm B$, fields are formed: this particular field configuration
is named the Glasma \cite{Lappi:2006fp} and it serves as the initial condition for the evolution of the classical gluon field after the collision
that can be studied by the Classical Yang-Mills (CYM) 
equations \cite{PhysRevC.89.024907,FUKUSHIMA2012108,PhysRevLett.111.232301,Berges:2020fwq,Gale:2012rq,
Gale:2012in,PhysRevD.97.076004} up to the formation of the quark-gluon plasma.

At the initial time $\bm E \cdot \bm B\neq 0$ in Glasma, that results in a non vanishing topological charge density, $\rho_T$: 
the initial condition then consists of approximately $N=\pi R_{A}^{2}Q_{s}^{2}$ uncorrelated 
domains of topological charge density \cite{KHARZEEV2002298}, where $R_A$ is the nucleus radius and $Q_s$ the saturation scale,
in which the amplitude of the fluctuations of the topological charge density is given by $\sqrt{N}\approx R_{A}Q_{s}$.  
It has been discussed that the high energy 
collisions might be described by the decay of instanton like configurations, the sphalerons 
\cite{KHARZEEV2002745,PhysRevC.75.044903,PhysRevD.67.014005}. 
The decay of these longitudinal electric and magnetic fields is in fact the decay of topological charges. 
Since $\rho_T$ is charge conjugation and parity ($\mathcal{CP}$) odd, 
it might be the source of large $\mathcal{CP}$ violating fluctuations in heavy ion collisions, see Refs.\cite{PhysRevLett.81.512,PhysRevD.61.111901,Schrempp:2004vj,KHARZEEV2008227}, and \cite{PhysRevD.97.034034,PhysRevC.82.057902,PhysRevD.98.014025,PhysRevD.98.071902,doi:10.1142/S0219887819501123,doi:10.1142/S2010194518600108} for recent reviews.

A nonzero $\rho_T$ naturally induces the production of chirality imbalance, $N_5$, by virtue of the chiral anomaly of 
Quantum Chromodynamics (QCD).
In this article, we focus on the topological charge density $\rho_T$ and the associated chiral charge density $N_5$
produced in the evolving Glasma. 
Differently from \cite{PhysRevC.89.024907,iida2014time} where the evolution has been followed up to late times, 
we will  focus on the very early stage, that is the proper time range in which the description based on CYM has some 
phenomenological interest for high energy nuclear collisions.  
In particular, we are interested to follow the evolution of the correlators of $\rho_T$ and of the chiral density
in the early stage, computing the size of the correlation domains as a function of the center of mass energy of 
the collision; we also compute the expectation value of the chiral density in the steady state, namely 
in the range or proper time in which the fields are diluted enough that further chiral density is not produced,
and study how this depends on the collision energy.
The latter estimate can have some phenomenological interest as it turns out that the proper time needed to form the 
steady state is in the same ballpark of the thermalization time, namely the time at which the CYM evolution 
is switched to that of the quark-gluon plasma via relativistic hydro or 
transport \cite{Ruggieri:2013bda,Ruggieri:2013ova,Ruggieri:2015tsa,Ruggieri:2015yea,Gale:2012rq,Gale:2012in,
Ryblewski:2013eja}:
the amount of chiral density that we compute will therefore be present at this stage
and it is likely to persist for the full evolution of the system up to hadronization time since
the relaxation time for $N_5$ in the hot quark-gluon plasma due to the helicity flipping processes turns out to be 
larger than the typical lifetime of the fireball produced in heavy ion collisions \cite{Manuel:2015zpa}.
This study is a natural continuation of previous works on the correlators in the evolving Glasma
\cite{PhysRevD.88.031503,DUMITRU20147,PhysRevC.79.024909}, 
and paves the way to future studies on the anomalous particle production by 
the chiral magnetic effect (CME) \cite{Fukushima_2019,Beni__2019,PhysRevB.93.155107,PhysRevD.78.074033,PhysRevD.86.071501,PhysRevD.82.054001}.

The article is organized as follows. In Sec.\ref{sec2}, we briefly review the McLerran-Venugopalan (MV) model and the classical Yang-Mills (CYM) equation. In Sec.\ref{sec3}, we introduce the topological charge density and chiral charge density via Adler-Bell-Jackiw anomaly equation, then we define the associated correlators. In Sec.\ref{sec4}, we show our results for topological charge density and chiral charge density correlations at different energy scale. Then, we estimate the size of the correlated domains for both quantities. At last, we analyze their time dependence as well as energy scale dependence. In Sec.\ref{sec5}, we draw our conclusion and make some outlooks.

\section{Glasma and classical Yang-Mills equation}\label{sec2}
In this section, we briefly review the McLerran-Venugopalan (MV) model, by which we describe the 
initial condition of the classical gluon field produced after the collision \cite{PhysRevD.49.2233,PhysRevD.49.3352,PhysRevD.50.2225,PhysRevD.54.5463}, which is then evolved by the classical Yang-Mills (CYM) equations . 
We remark that in our notation the gauge fields have been rescaled by the QCD coupling $A_{\mu}\rightarrow A_{\mu}/g$
therefore $g$ does not appear explicitly in the equations.

\subsection{Glasma}\label{sec2.1}
In the MV model, the color charge densities $\rho^a$ act as  static sources of the transverse CGC fields 
in two colliding nuclei: 
they are assumed to be random variables that, for each of the two colliding nuclei, 
are normally distributed with zero average and variance specified by the equation
\begin{equation}
	\langle\rho^a(\mathbf{x}_{\perp},\eta_1)\rho^b(\mathbf{y}_{\perp},\eta_2)\rangle=
	(g^2\mu)^2\delta^{ab}\mathbf{\delta}(\mathbf{x}_{\perp}-\mathbf{y}_{\perp})
	\delta(\eta_1-\eta_2),\label{eq1}
\end{equation}
where $a$ and $b$ denote the adjoint color indices, $\mathbf{x}_{\perp}$ and $\mathbf{y}_{\perp}$
denote transverse plane coordinates and $\eta_1$, $\eta_2$ the space time rapidities. 
In this work, we limit ourselves to a SU(2) glasma, therefore, $a,b=1,2,3$. 
In the MV model, $g^2\mu$ is the only energy scale which is related to the saturation momentum $Q_s$: 
rather, we simply refer to the estimate of \cite{LappiWilson} namely that $Q_s/g^2\mu\approx 0.57$. 
Due to the $\delta-$functions in Eq.~\eqref{eq1} the color charge densities 
are uncorrelated in transverse plane as well as in rapidity.  
To specify the initial condition, namely the Glasma fields,
it is convenient to work in Bjorken coordinates $(\tau,\eta)$ in the radial gauge, where
\begin{eqnarray}
    &\sqrt{2}x^{\pm}=\tau e^{\pm\eta},\label{eq2}\\
	&A_{\tau}=x^{+}A^{-}+x^{-}A^{+}=0.\label{eq3}
\end{eqnarray}
In order to compute Glasma fields we firstly solve the Poisson equations, namely
\begin{eqnarray}
	-\mathbf{\nabla}\cdot\alpha^{(A)}(\mathbf{x}_{\perp})=\rho^{(A)}(\mathbf{x}_{\perp}),\label{eq4}\\
	-\mathbf{\nabla}\cdot\alpha^{(B)}(\mathbf{x}_{\perp})=\rho^{(B)}(\mathbf{x}_{\perp}),\label{eq5}
\end{eqnarray}
where $A$ and $B$ denote the two colliding nuclei. The solutions of these equations are 
\begin{eqnarray}
\alpha^{(A)}_{i}(\mathbf{x}_{\perp})=iU^{(A)}(x_{\perp})\partial_iU^{(A)\dagger}(x_{\perp}),\label{eq6}\\
\alpha^{(B)}_{i}(\mathbf{x}_{\perp})=iU^{(B)}(x_{\perp})\partial_iU^{(B)\dagger}(x_{\perp}).\label{eq7}
\end{eqnarray}
while the Wilson line is defined as $U(x_{\perp})\equiv\mathcal{P}{\rm exp}\left(-i\int d\mathbf{z}^{\mu}\alpha^{\mu}(\mathbf{z(x_{\perp})})\right)$, with $\mathcal{P}$ is the path order operator and $\mathbf{z}(x_{\perp})$ is a trajectory. In terms of these fields, the glasma gauge potential at $\tau=0^+$ can be written as \cite{PhysRevD.52.6231,PhysRevD.52.3809}:
\begin{eqnarray}
	&A_{i}=\alpha_{i}^{(A)}+\alpha_{i}^{(B)}, i=x,y,\label{eq8}\\
	&A_{\eta}=0,\label{eq9}
\end{eqnarray}
Solving the Yang-Mills equation near the light cone, one finds that the transverse color electric and color magnetic fields vanish as $\tau\rightarrow 0$, but the longitudinal electric and magnetic fields are non-vanishing \cite{Fukushima_2007}:
\begin{eqnarray}
	&E_{\eta}=i\sum_{i}[\alpha_{i}^{(A)},\alpha_{i}^{(B)}], \label{eq10}\\
	&B_{\eta}=i([\alpha_{x}^{(A)},\alpha_{y}^{(B)}]+[\alpha_{x}^{(B)},\alpha_{y}^{(A)}]).\label{eq11}
\end{eqnarray}
In all the discussion above we have neglected the possibility of fluctuations that, among other things,
would break the longitudinal  boost invariance:
while these are relevant for the onset of the hydrodynamical flow
\cite{Dusling:2010rm,Romatschke:2006nk,Romatschke:2005ag,Romatschke:2005pm,PhysRevLett.111.232301,
PhysRevD.97.076004,FUKUSHIMA2012108,PhysRevC.89.024907}, 
they are not crucial for the production of chiral density and for the evolution
of the topological charge density; in any case, we plan to add these fluctuations in near future works.
We also assume that $g^2\mu$ is the same in the two colliding nuclei and it has no dependence on the transverse plane
coordinates: we will remove this assumption in the future, to mimic the energy density profile that would be produced
in realistic collisions \cite{Gale:2012rq,Gale:2012in}.

\subsection{The classical Yang-Mills equation}\label{sec2.2}
After preparing the initial condition of CYM equations, we solve these numerically. 
In the gauge $A_{\tau}=0$, the Lagrangian density reads
\begin{equation}
	\mathcal{L}=\frac{1}{2}{\rm Tr}\big[-\frac{2}{\tau^2}(\partial_{\tau}A_{\eta})^2-2(\partial_{
	\tau}A_{i})^{2}+\frac{2}{\tau^2}F^{2}_{\eta i}+F_{ij}^{2}\big],\label{eq12}
\end{equation}
and the canonical momenta are given by
\begin{eqnarray}
	E_{i}=\tau\partial_{\tau}A_{i},\label{eq13}\\
	E_{\eta}=\frac{1}{\tau}\partial_{\tau}A_{\eta}.\label{eq14}
\end{eqnarray}
As a consequence, the Hamiltonian density is
\begin{equation}
	\mathcal{H}={\rm Tr}\big[\frac{1}{\tau^2}E_{i}^2+E_{\eta}^2+\frac{1}{\tau^2}F_{\eta i}^2+\frac{1}{2}F_{ij}^2\big].\label{eq15}
\end{equation}
The CYM equations in Bjorken coordinates are
\begin{eqnarray}
	&\partial_{\tau}E_{i}=\frac{1}{\tau}\mathcal{D}_{\eta}F_{\eta i}+\tau\mathcal{D}_{j}F_{ji},\label{eq16}\\
	&\partial_{\tau}E_{\eta}=\frac{1}{\tau}\mathcal{D}_{j}F_{j\eta},\label{eq17}
\end{eqnarray}
where $\mathcal{D}_{\mu}=\partial_{\mu}+iA_{\mu}$ is covariant derivative. Besides, 
we define $B_i=-\epsilon^{ij}F_{j\eta}$ and $B_{\eta}=-\frac{1}{2}\epsilon^{ij}F_{ij}$
as the $i$ and $\eta$ components of the color magnetic field. 
In the next section, these equations will be solved in a $4~\mathrm{fm}\times 4~\mathrm{fm}$ box with 
periodic boundary conditions on the transverse plane. The lattice spacing that we use is $a=0.04$ fm. 
We solve the CYM equations via the fourth order Runge-Kutta method as in \cite{PhysRevD.97.076004,iida2014time}: 
results with the given approach are in good agreement with the Yang-Mills solver
based on the gauge links \cite{FUKUSHIMA2012108,PhysRevLett.111.232301}, and we leave the more rigorous implementation of the numerical problem based on gauge links to a future work.

We have checked that within our Yang-Mills solver the violation of   the Gauss's law, $D_a E^a=0$, is minimal
within $\tau\leq 0.3$ fm/c, and is mitigated by lowering the lattice spacing.
The small violation of Gauss's law is harmless for everything that we compute in this work.
In particular, we have found no concrete trace of spurious sources of the gluon fields in the evolution of the energy density, $\varepsilon$,
since we are able to recover the free streaming behavior 
$\varepsilon\propto 1/\tau$ within $g^2\mu\tau\lesssim 1$
in agreement with~\cite{FUKUSHIMA2012108} where Yang-Mills equations have been solved
using the formulation based on the gauge links. We have also verified by random gauge rotations that 
 as long as we compute gauge-invariant
quantities, for example $\mathrm{Tr}(\bm E \cdot \bm B)$, our formulation produces gauge invariant results
within numerical uncertainties.

\section{The topological charge and the chiral densities}\label{sec3}

As mentioned above, immediately after the collision nonzero longitudinal components of the color electric, $\bm E = E_a T_a$, 
and color magnetic, $\bm B = B_a T_a$  fields are formed (here and in the following we use boldface to denote 
vectors in color space): 
because of the Adler-Bell-Jackiw anomaly equation \cite{PhysRev.177.2426,J1969A}, these lead to the
nonconservation of the $U(1)$ axial symmetry of QCD, that in Bjorken coordinates is written as  
\begin{equation}
\left(\partial_\tau + \frac{1}{\tau}\right)j_5^\tau
=  \frac{1}{8\pi^2}{\rm{Tr}}\vec{\bm E}\cdot \vec{\bm B},
\label{eq:chan1}
\end{equation}
where, in the same coordinate system,
\begin{equation}
\vec{\bm E} \cdot \vec{\bm B} = \frac{{\bm E}_x {\bm B}_x + {\bm E}_y {\bm B}_y } {\tau^2} + {\bm E}_\eta {\bm B}_\eta. 
\end{equation}
In writing Eq.~\eqref{eq:chan1} we have used the boost-invariance assumption as well as that the divergence of the axial current in the
transverse plane vanishes, the latter requirement implying that there is no net flux of the axial current across the transverse plane.
The gauge invariant quantity on the right hand side of Eq.~\eqref{eq:chan1} is called the topological charge density since
its integral over the full volume and on the time history of the system gives the change in the Chern-Simons number of the
gluon configuration,
\begin{equation}
\Delta N_\mathrm{CS} = \frac{1}{8\pi^2}\int d^4x{\rm Tr}\vec{\bm E}\cdot\vec{\bm B}.
\end{equation}
Therefore, here we call the invariant on the right hand side of Eq.~\eqref{eq:chan1} as the topological charge density,
\begin{equation}
\rho_T(\tau,x_\perp) = \frac{1}{8\pi^2}{\rm Tr}\vec{\bm E}\cdot \vec{\bm B},\label{eq:EdotB}
\end{equation}
where we have made explicit the dependence of this quantity on $\tau$ and $x_\perp$ that comes from that of $\bm E$ and $\bm B$.

It is possible to use $j_5^\tau$ to define a chiral charge per unit volume, namely
\begin{equation}
j_5^\tau(\tau,x_\perp) = \frac{dN_5}{\tau d\eta d^2x_\perp},\label{eq:def1}
\end{equation}
where $N_5$ corresponds to the chiral charge and $\tau d\eta d^2x_\perp$ is the 3-dimensional volume of a cell
with extension $d^2x_\perp$ in the transverse plane and $\tau d\eta$ in the longitudinal direction; from this we can define
the chiral charge distribution in the transverse plane per unit rapidity as
\begin{equation}
n_5(\tau,x_\perp)\equiv\frac{dN_5}{d\eta d^2x_\perp} = \tau j_5^\tau(\tau,x_\perp).\label{eq:chde1}
\end{equation}
In this study, we solve the CYM equations to get $\rho_T$ by means of Eq.~\eqref{eq:EdotB}
and we use it in the right hand side of  Eq.~\eqref{eq:chan1} that can be formulated in terms of $n_5$ by writing $j_5^\tau=n_5/\tau$
then solving for $n_5$, namely
\begin{equation}
 n_5(\tau,x_\perp)=\int_0^\tau d\tau^\prime \rho_T(\tau^\prime,x_\perp) \tau^\prime.\label{eq:forn5exact}
\end{equation}

Here we study how topological charge density and chiral density correlation domains form with time,
and we estimate their size on the transverse plane. According to Eq.~\eqref{eq:chde1} quarks should form as soon as the chiral anomaly starts to play its game: in principle, these quarks bring additional sources in the Yang-Mills equations~\eqref{eq16} and~\eqref{eq17}, but we neglect these because we limit ourselves to study the 
early stages after the collision, and in this stage the number of quarks per unit volume is small leaving the dynamics dominated by the gluons.

 Next we turn to the definition of the gauge invariant correlators that we are interested to;
in this study, we consider only equal time correlators, leaving the different times ones  to a future study.
In continuum limit, the correlator of $\rho_T$ we analyze is defined as
\begin{equation}
	C(|x_{\perp}|)=\langle\rho_T(0)\rho_T(|x_{\perp}|)\rangle,
	\label{eq:cc11}
\end{equation}
where we have made explicit the dependence on the transverse plane coordinate, $x_\perp$ and $\langle\rangle$ denotes
the ensemble average. 
On the lattice, this average is done by running a finite amount of events, $N_\mathrm{events}$, 
then summing over all these and dividing by $N_\mathrm{events}$, where each event is initialized
with a different random condition. 
In addition to this average, we introduce an average over the volume of the box in order
to improve the statistics. Therefore, we compute the average of an observable $\mathcal{O}$ as
\begin{equation}
	\langle\mathcal{O}\rangle=\frac{1}{N_\mathrm{events}}\sum_\mathrm{events}\frac{1}{N}\sum_{x_{j}} \mathcal{O}(x_{j}),
\end{equation}
where  
$x_{j}$ denotes the ${\rm j}$th lattice point and N is the number of lattice points involved in the summation: 
concretely, for a given $x_{j}$ we sum the  $\mathcal{O}(x_{j})$ in the four directions in the transverse plane, namely
we add up the values of $\mathcal{O}$ computed at the points $(x_i,0)$, $(-x_i,0)$, $(0,y_i)$ and $(0,-y_i)$.

Similarly to Eq.~\eqref{eq:cc11} we define the correlator of the chiral density,
\begin{equation}
	D(|x_{\perp}|)=\langle n_5(0) n_5(|x_{\perp}|)\rangle.
	\label{eq:cc11a}
\end{equation}
From this correlator we can also define $\langle n_5^2\rangle \equiv  \langle n_5(|x_{\perp}|) n_5(|x_{\perp}|)\rangle$ 
that estimates the amount of fluctuations of chiral density in the transverse plane and that has phenomenological interest like in Refs.\cite{PhysRevC.82.057902,PhysRevD.98.071902}.
The study of the correlators~\eqref{eq:cc11a} and~\eqref{eq:cc11} is useful because these allow to 
estimate the size of the correlation domains of chiral and topological densities in the early stage
of high energy nuclear collisions, as we will discuss in the next section.

\section{Results}\label{sec4}

In this section we present our results: we firstly discuss the correlation of the topological charge, then we turn to that 
of the chiral density. Finally, we estimate the size of the correlated domains for both quantities. 
In particular, we analyze the time dependence as well as the dependence on the energy scale $g^2\mu$
of the correlators.
 
\subsection{Correlation of the topological charge density}\label{sec4.1}

\begin{figure}[t!]
\begin{center}
\includegraphics[width=\linewidth]{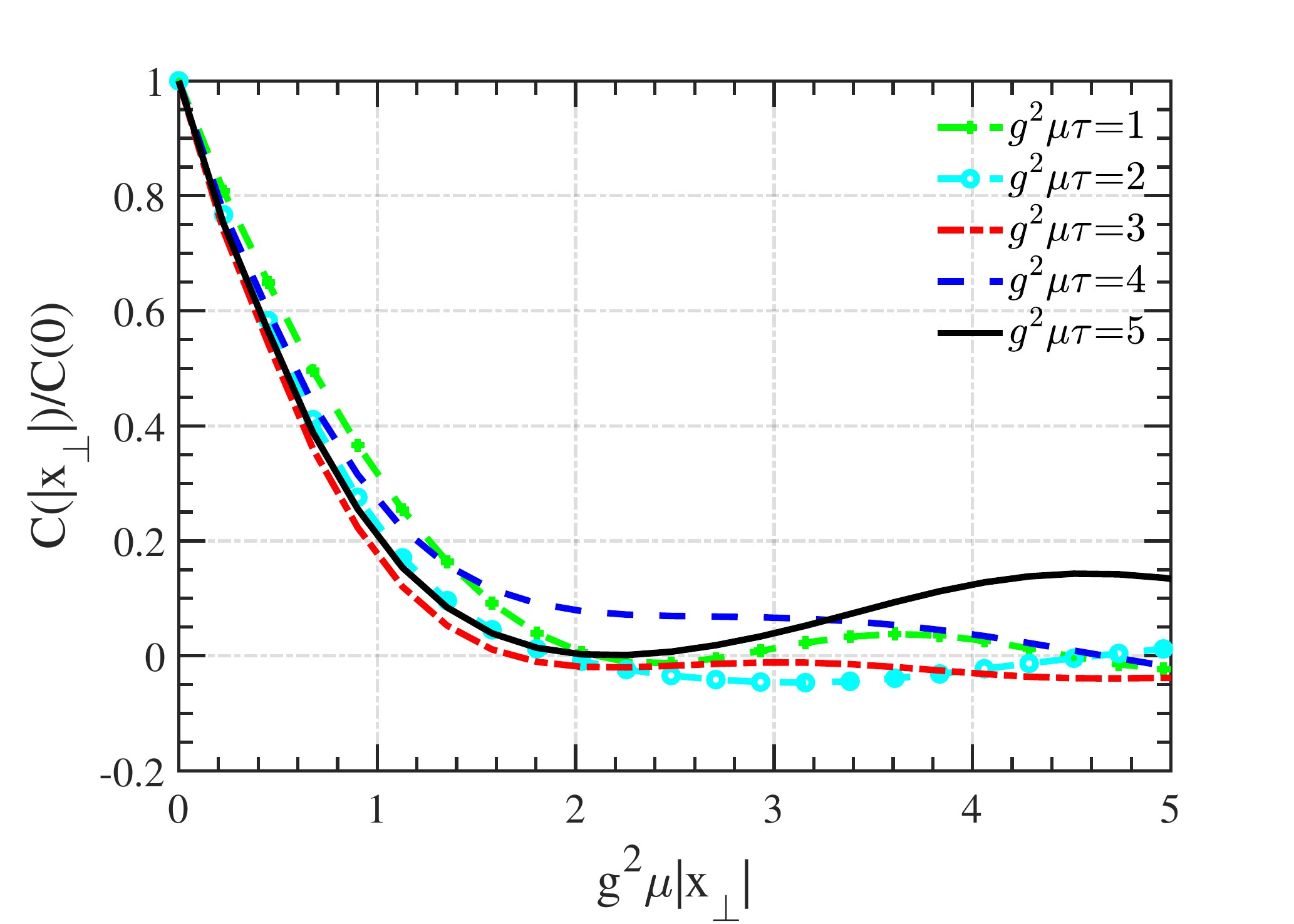}
\end{center}
\caption{ Correlator of the topological charge density $\rho_T$ as a function of transverse plane coordinate at different values
of the proper time. We have put $g^2\mu=1$ GeV and the lattice spacing $a=0.04$ fm giving $g^2\mu a=0.2$. 
All the correlators are normalized by 
	their value at $x_\perp=0$.}  
\label{fig1}
\end{figure}

In Fig. \ref{fig1}, we plot the correlator of the topological charge density, Eq.~\eqref{eq:cc11},
versus the transverse plane coordinate at different proper times;
the correlators have been normalized to their values at $x_\perp=0$ 
(an overall constant in each correlator is not relevant at all if we focus
on the correlation length), and have been computed
for $g^{2}\mu=1$ GeV. 
The calculations have been performed on a lattice with transverse size $4~\mathrm{fm}\times 4~\mathrm{fm}$
with a lattice spacing $a=0.04$ fm giving $g^2\mu a = 0.2$. 
The ensemble average has been performed by averaging over $N_\mathrm{events}=200$ events,
and we have checked that this $N_\mathrm{events}$ is enough to get numerical convergence.

The results in Fig. \ref{fig1} show that the correlation of $\rho_T$ decays quickly in transverse plane: 
in fact, for all the cases shown in the figure
already for $x=0.16$ fm, corresponding to a dimensionless distance of $g^2\mu x=0.8$, correlation reduces
to approximately the $40\%$ of the initial value. 
Although the qualitative behavior of the correlators is the same at each of the finite time we
analyze, we notice some fluctuation of the shape as time evolution goes on: compare
for example the cases at $g^2\mu\tau=2$, $3$ and $4$ in Fig.~\ref{fig1}: 
these fluctuations are due to the continuous exchange of energy between the longitudinal and the
transverse fields that happens in the evolution, but they lead to almost no quantitative effect on the
correlation length, see below. 
We also notice that some modest anticorrelation shows up in time: an anticorrelation was also
obtained for the color electric and color magnetic fields \cite{DUMITRU20147,PhysRevD.97.076004} and in this case shows the 
tendency of the 
topological charge density to flip its sign on length scales much larger than the correlation domains.  

\begin{figure}[t!]
	\centering
	\includegraphics[width=\linewidth]{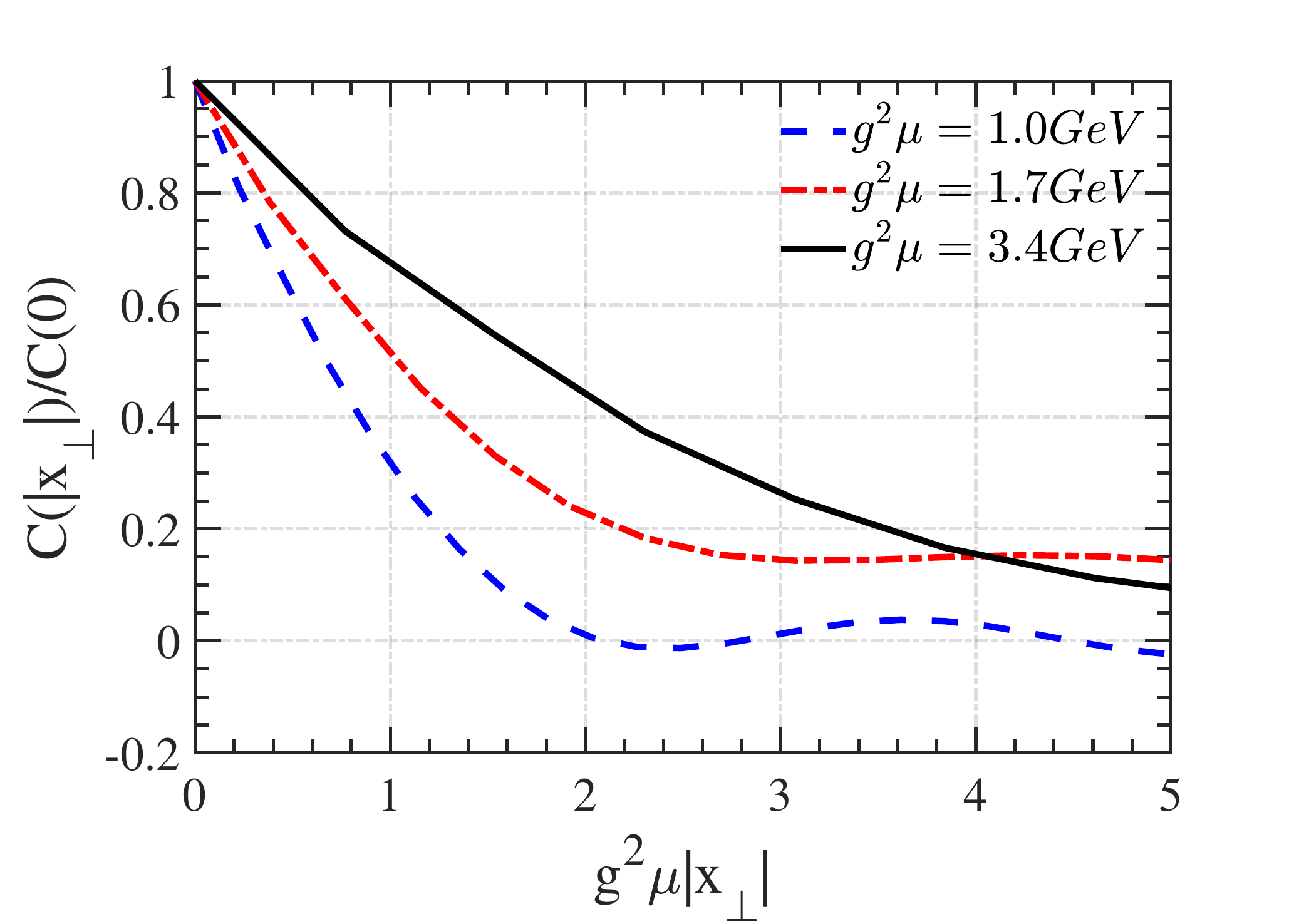}\\
	\includegraphics[width=\linewidth]{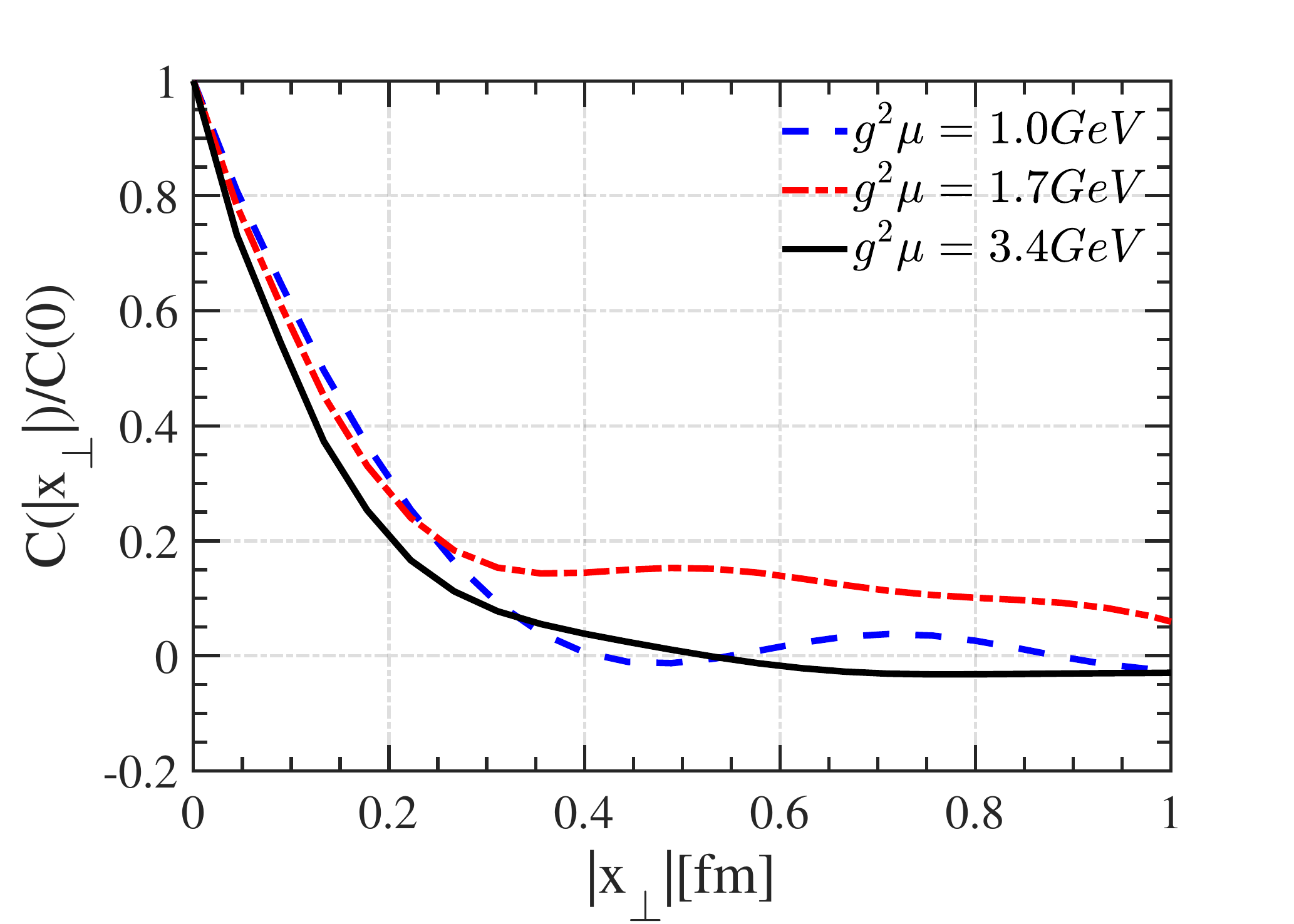}
	\caption{ Correlator of the topological charge density $\rho_T$ as a function of transverse plane coordinate at different saturation energies. The lattice parameters are the same as in Fig.\ref{fig1}.
In the upper panel we plot the correlators as a function of the dimensionless length $g^2\mu |x_\perp|$, while in the lower panel
we have the same quantities versus the physical length $|x_\perp|$.	
	 All the correlators are normalized by 
	their value at $x_\perp=0$, and computed at $g^2\mu\tau=1$.}
		\label{fig2}
\end{figure}

In Fig. \ref{fig2}, we plot the correlator of the topological charge density versus the transverse plane coordinate for
three values of $g^2\mu$: this piece of information never appeared in the literature before,
but it is important because changing $g^2\mu$ amounts to change
the energy of the collision and a larger $g^2\mu$ corresponds to a larger collision energy. In this paper, we have worked up to $g^2\mu=3.4$ GeV that corresponds roughly to the estimate
for this quantity for a Au-Au collision at the maximum RHIC energy. 
We find that increasing the $g^2\mu$ results in a broader correlator when this is studied as a function of the 
dimensionless length $g^2\mu|x_\perp|$. 
In the lower panel of Fig. \ref{fig2} we plot the same quantity versus the physical length:
increasing the $g^2\mu$ results in a quicker decay of the correlator, which suggests that the correlation length of the
topological charge density becomes smaller.

\begin{figure}[t!]
	\centering
	\includegraphics[width=\linewidth]{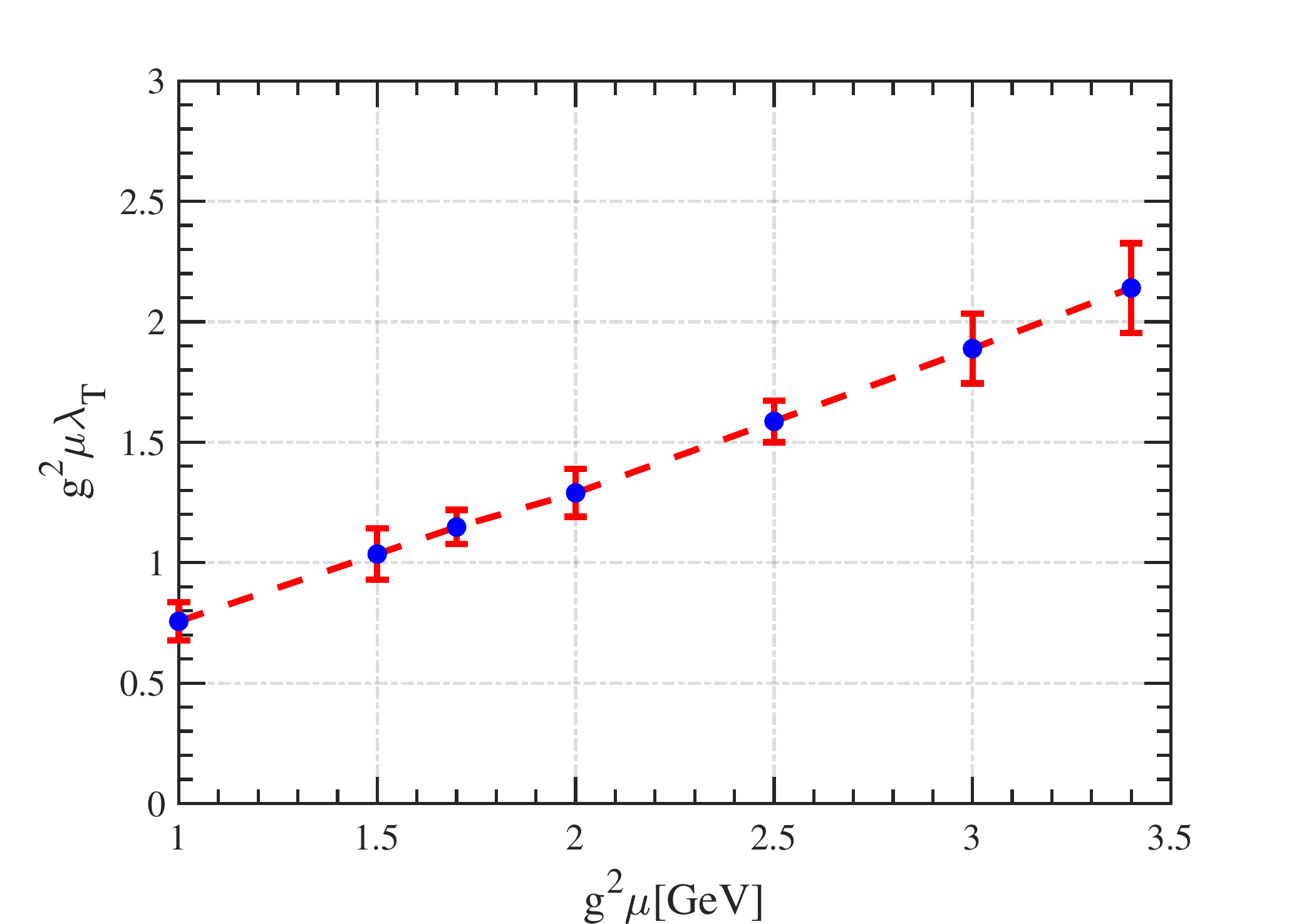}\\
	\includegraphics[width=\linewidth]{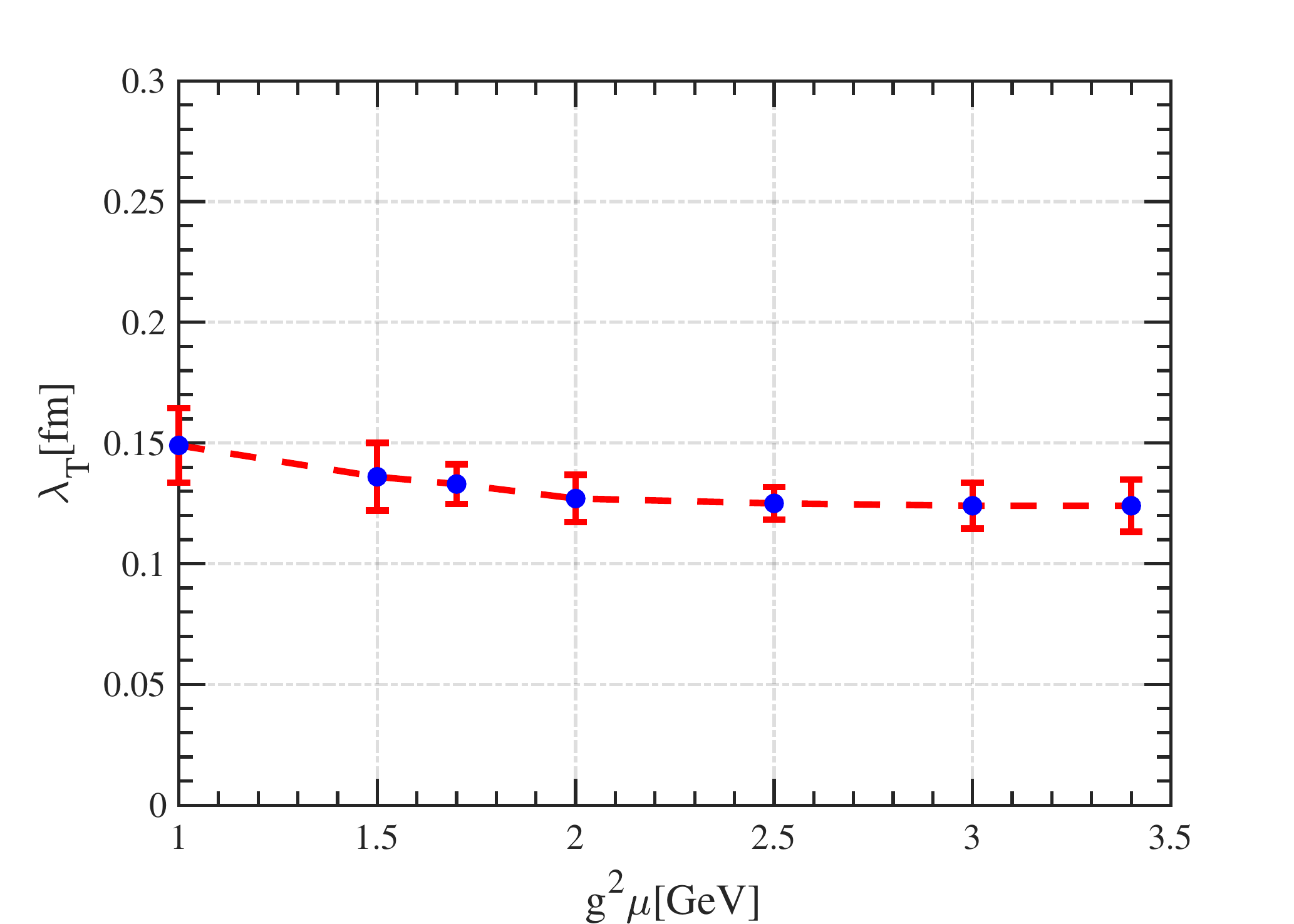}
	\caption{{\it Upper panel.} Correlation length of the topological charge density $\rho_T$, in units of $g^2\mu$, 
	versus  $g^2\mu$. {\it Lower panel}.  Physical correlation length of $\rho_T$ versus $g^2\mu$.
	The parameters are the same as in Fig. \ref{fig1}.}
	\label{fig4}
\end{figure}

The results in Fig. \ref{fig2} allow to compute a correlation length, $\lambda_T$, of $\rho_T$ in the transverse plane: 
we define this length by the value of $|x_\perp|$ such that the correlator decays to $1/e$ of its initial value.
Since the correlators depend on time, this definition in principle depends on time as well: therefore, we perform an average
over the full time history of the system in addition to the ensemble average.
Concretely, what we do is that firstly we compute the correlators at any time by taking the ensemble average of the
proper quantity; then, for each time we find the value of $|x_\perp|$ such that $C(x_\perp)/C(0)=1/e$,
and we average over these values, keeping also the maximum and the minimum of the range to define the uncertainty.
We show the result of this definition in Fig.~\ref{fig4} in which we plot the correlation length versus $g^2\mu$.
In the upper panel we plot the dimensionless correlation length, while in the lower panel we show the
correlation length in units of fm.
In the figures, the blue dots correspond to the average value while the error bar 
denotes the maximum and the minimum value of $\lambda_T$ achieved during time evolution.
We find that  on average $g^2\mu \lambda_T$ lies within about $0.5-2.5$ in the range of $g^2\mu$ analyzed, 
which is consistent with the expectation
that the correlation domains of $\rho_T$ are of the same transverse size of the correlation domains of color electric and color magnetic
fields, see for example \cite{PhysRevD.97.076004}. 
Changing to physical units, we find that $\lambda_T$ lies in the range $(0.1~\mathrm{fm},0.2~\mathrm{fm})$, 
namely the domains remain microscopic with respect to the transverse size of nucleons. Altogether, these results show that increasing the energy of the collision,
the number of correlation domains of $\rho_T$ in the transverse plane increases as $\approx S/\lambda_T^2\approx S (g^2\mu)^2$
where $S$ denotes the transverse area of the nucleon, so the density of these domains in the transverse plane
behaves as $\approx (g^2\mu)^2$: increasing the energy results in more and finer domains of topological charge density
in the transverse plane.

\subsection{Production of chiral density}\label{sec4.2}

\begin{figure}[t!]
	\centering
	\includegraphics[width=\linewidth]{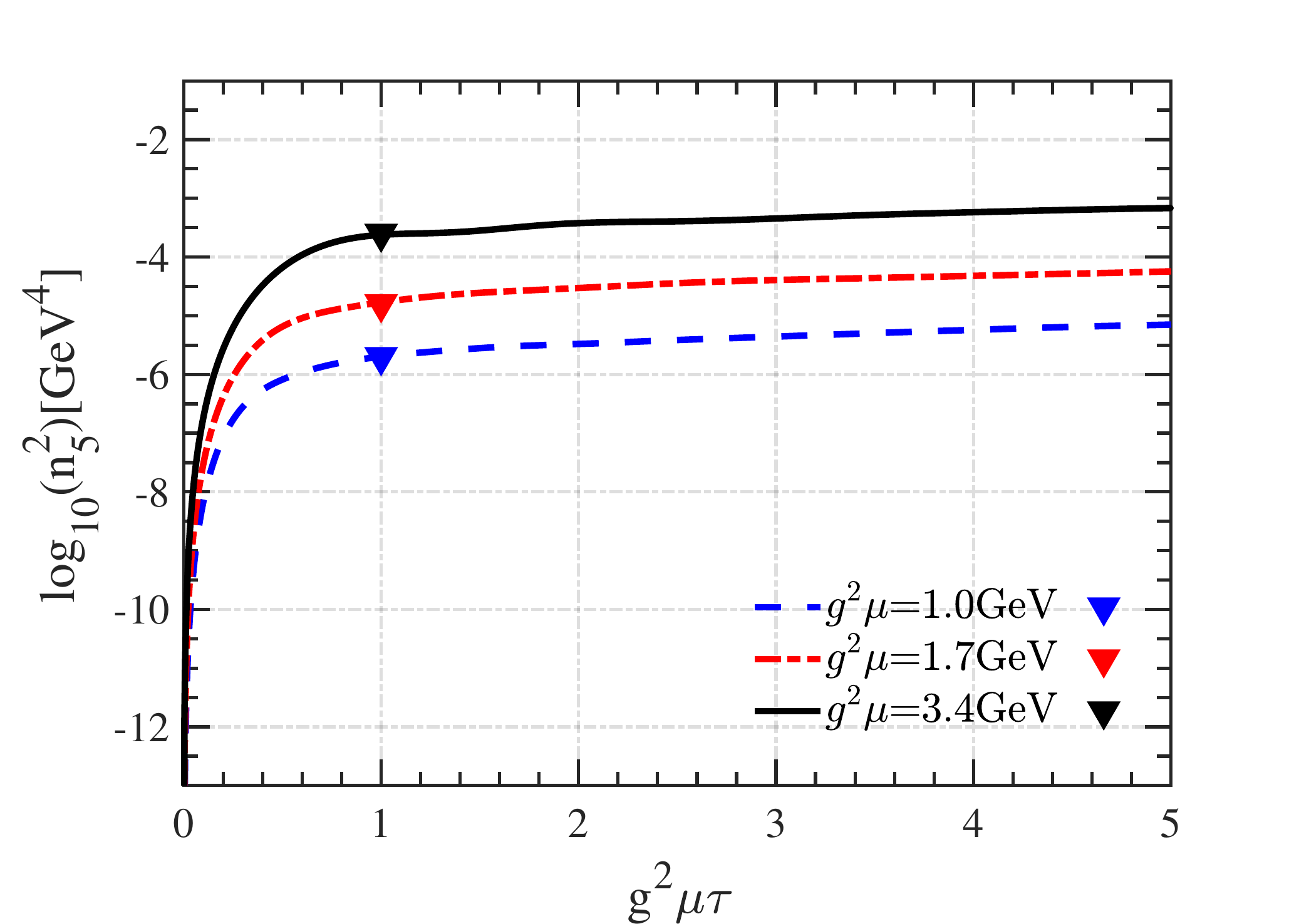}\\
	\includegraphics[width=\linewidth]{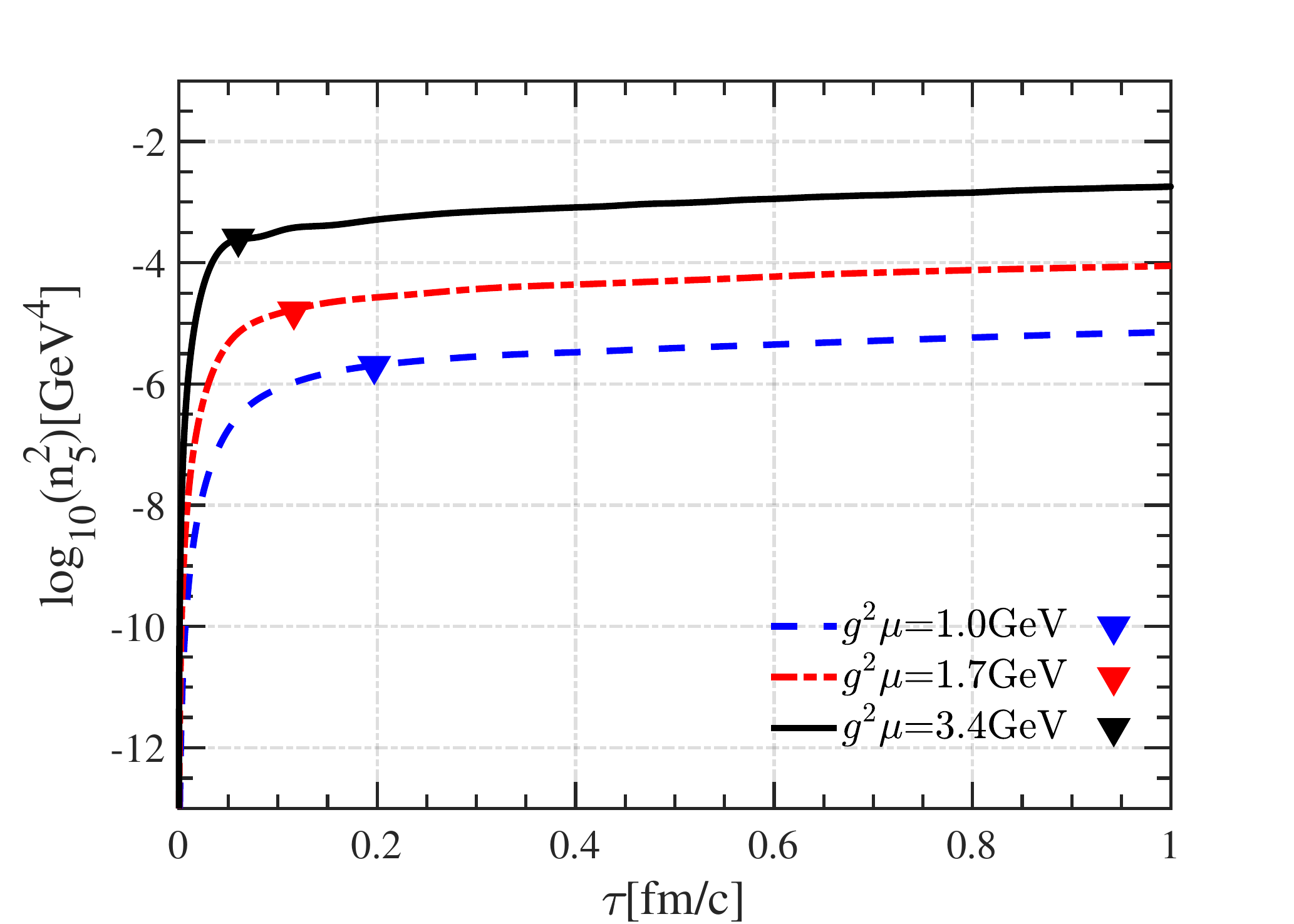}
	\caption{ {\it Upper panel.} Squared averaged chiral density, shown in a log-scale,  versus proper time
measured in units of $g^2\mu$, for three values of $g^2\mu$. {\it Lower panel.} The same quantity plot versus the time in fm/c.
Lattice setup corresponds to that of Fig.~\ref{fig1}.}
	\label{Fig:chde}
\end{figure}

In Fig.~\ref{Fig:chde} we plot the averaged $n_5^2 $ , shown in a log-scale, versus proper time for three values of $g^2\mu$.
In the upper panel we show the time measured in units of $g^2\mu$ while in the lower panel
we plot the quantity versus time measured in fm/c.
We notice that in all the cases considered here, for $g^2\mu\tau\approx 1.0$ the bulk of chiral density
is already formed, since a steady state is reached within this time range: this fixed point is due to the dilution 
of the fields with the expansion that lowers the value of the topological charge density and stops the production of $n_5$
via the chiral anomaly.
The actual value of the average $n_5^2$ in the steady state depends on $g^2\mu$, which is obvious
because increasing the latter results in a higher average of the gluon fields in the initial condition.

In both panels of Fig.~\ref{Fig:chde} we have highlighted by filled triangles the values of the chiral 
charge density at $g^2\mu\tau=1$ that corresponds roughly to the proper time range necessary to reach the steady state;
we denote this time by $\tau_\mathrm{s}$.
In calculations based on CYM the $g^2\mu$ is  a parameter:
it is interesting to give numerical estimates of the chiral charge density 
in the steady state and of its dependence on $g^2\mu$.
From Fig.~\ref{Fig:chde} we calculate the squared chiral charge per unit volume and unit rapidity
at $\tau=\tau_s$,
\begin{equation}
\frac{\langle N_5^2\rangle}{V^2} \equiv \frac{\langle n_5^2 \rangle}{\tau_\mathrm{s}^2},\label{eq:definition_ooo}
\end{equation}
where the physical volume per unit rapidity is $V=A_T\times\tau_\mathrm{s}$ with $A_T$ denoting the transverse area 
and $\tau_\mathrm{s}=1/g^2\mu$,
and $\langle\rangle$ denotes ensemble and transverse plane average
[we remind that our definition of $n_5$ corresponds to chiral charge per unit transverse area and unit rapidity, see Eqs.~(23) and~(24)].
Our findings in the steady state are 
$\sqrt{\langle N_5^2\rangle/V^2}\approx 1.79\times 10^{-1}~\rm fm^{-3}$ at $g^2\mu= 1.0~\rm GeV$, 
$\sqrt{\langle N_5^2\rangle/V^2}\approx 8.85\times 10^{-1}~\rm fm^{-3}$ at $g^2\mu= 1.7~\rm GeV$,
$\sqrt{\langle N_5^2\rangle/V^2}\approx 6.48~\rm fm^{-3}$ at $g^2\mu= 3.4~\rm GeV$.

\subsection{Correlation of the chiral density}\label{sec4.3}

\begin{figure}[t!]
	\centering
	\includegraphics[width=\linewidth]{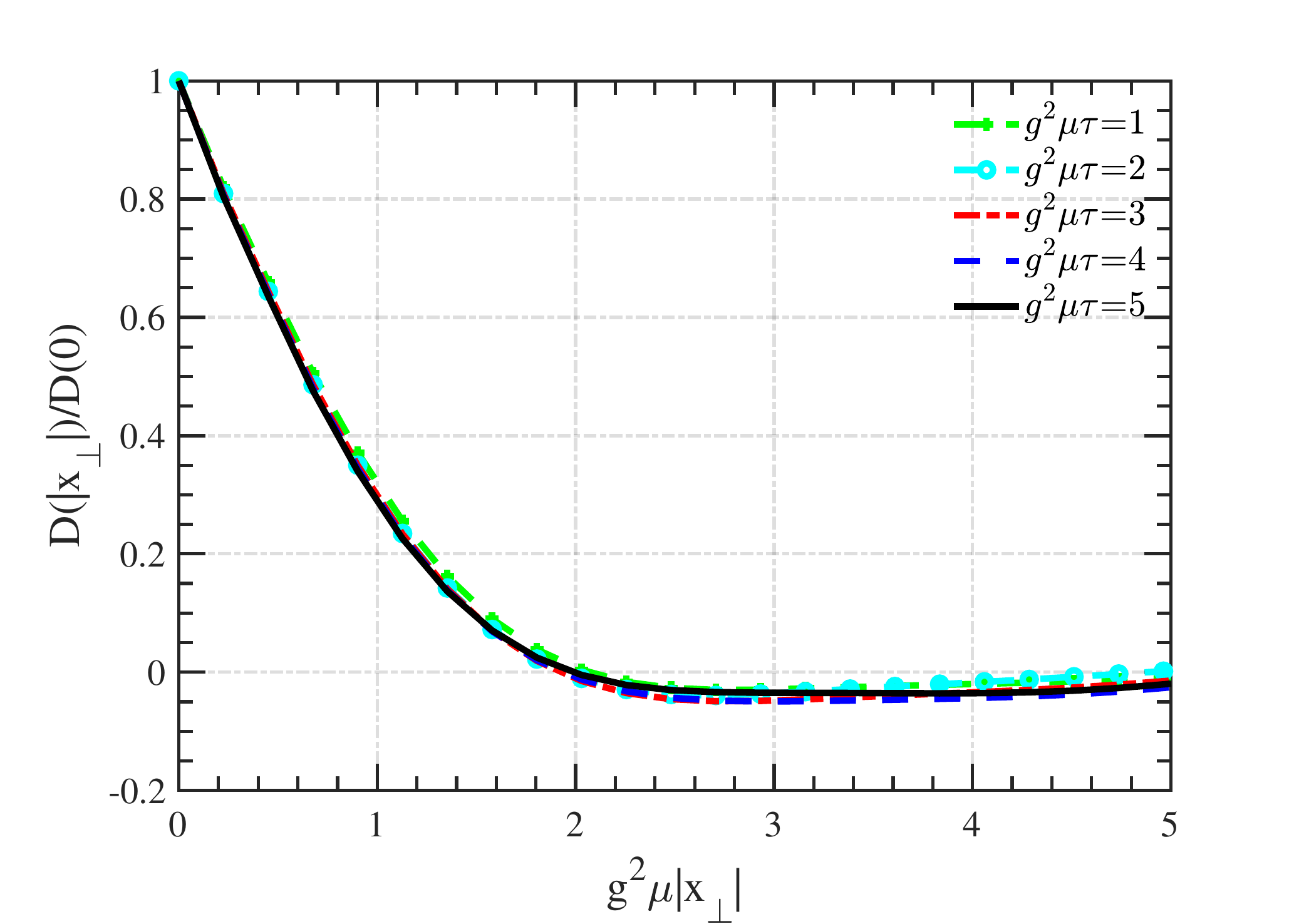}\\
	\includegraphics[width=\linewidth]{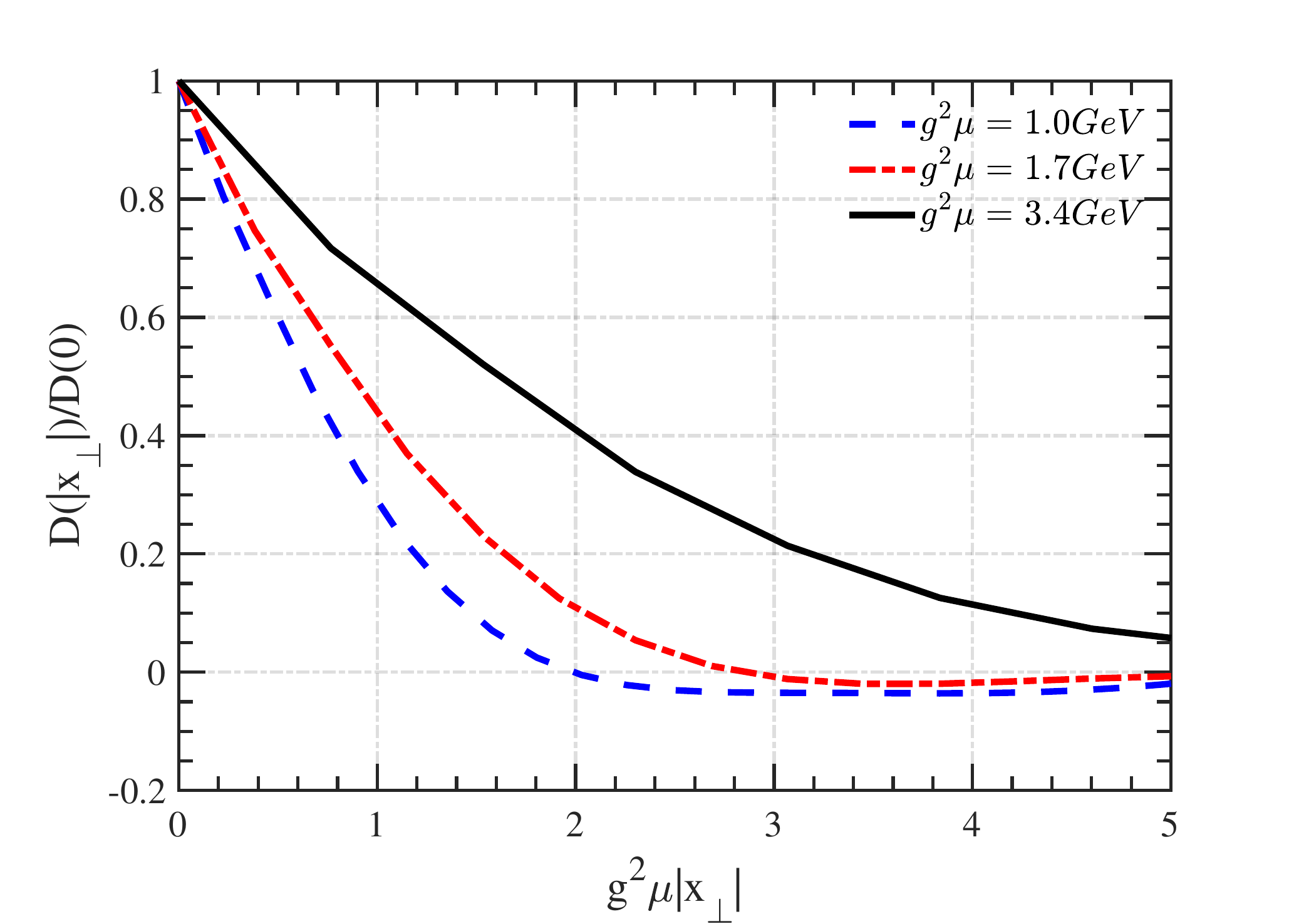}
	\caption{{\it Upper panel}.
	Correlator of the chiral charge density $n_5$ as a function of transverse plane coordinate at different values of the proper time. We have put $g^2\mu=1$ GeV and the lattice setup corresponds to that of Fig.~\ref{fig1}.
	{\it Lower panel}. Correlator of $n_5$ versus $g^2\mu|x_\perp|$ for three values of $g^2\mu$, computed at $g^2\mu\tau=1$. 
	Lattice setup corresponds to that of Fig.~\ref{fig1}.}
	\label{fig5}
\end{figure}

In the upper panel of 
Fig.~\ref{fig5} we plot the correlator of the chiral density (normalized to $x_\perp=0$ in each case) versus $g^2\mu|x_\perp|$,
for several values of the proper time, for $g^2\mu=1$ GeV.
Increasing time results in a progressive broadening of the correlator.
In the lower panel of Fig.~\ref{fig5} we plot $D(x_\perp)/D(0)$ at fixed time $g^2\mu\tau=1$ for three different values of $g^2\mu$.
In agreement with the results for $\rho_T$ presented in the previous subsection, 
increasing $g^2\mu$ affects mildly the correlator of $n_5$.

\begin{figure}[t!]
		\centering
		\includegraphics[width=\linewidth]{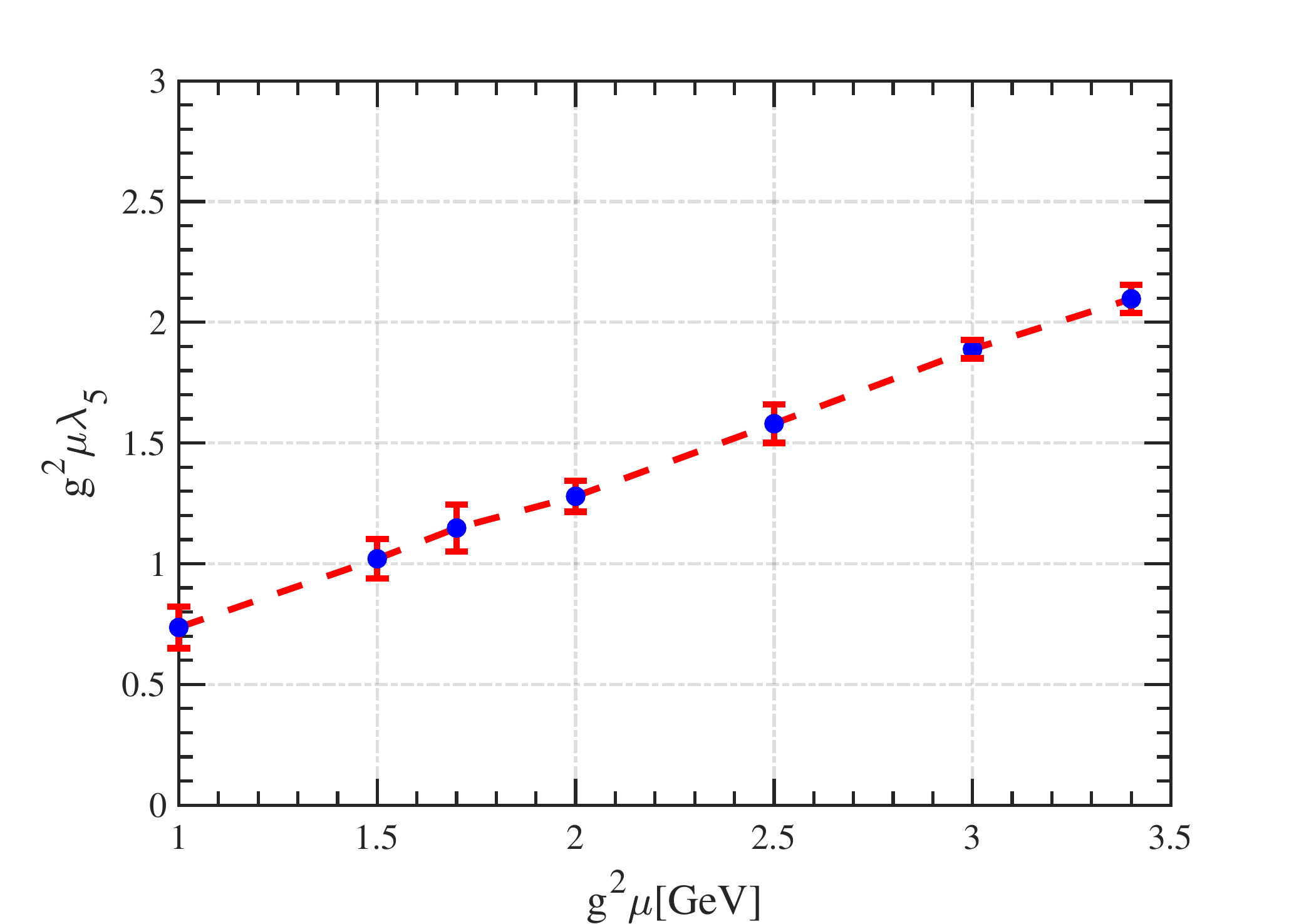}\\
		\includegraphics[width=\linewidth]{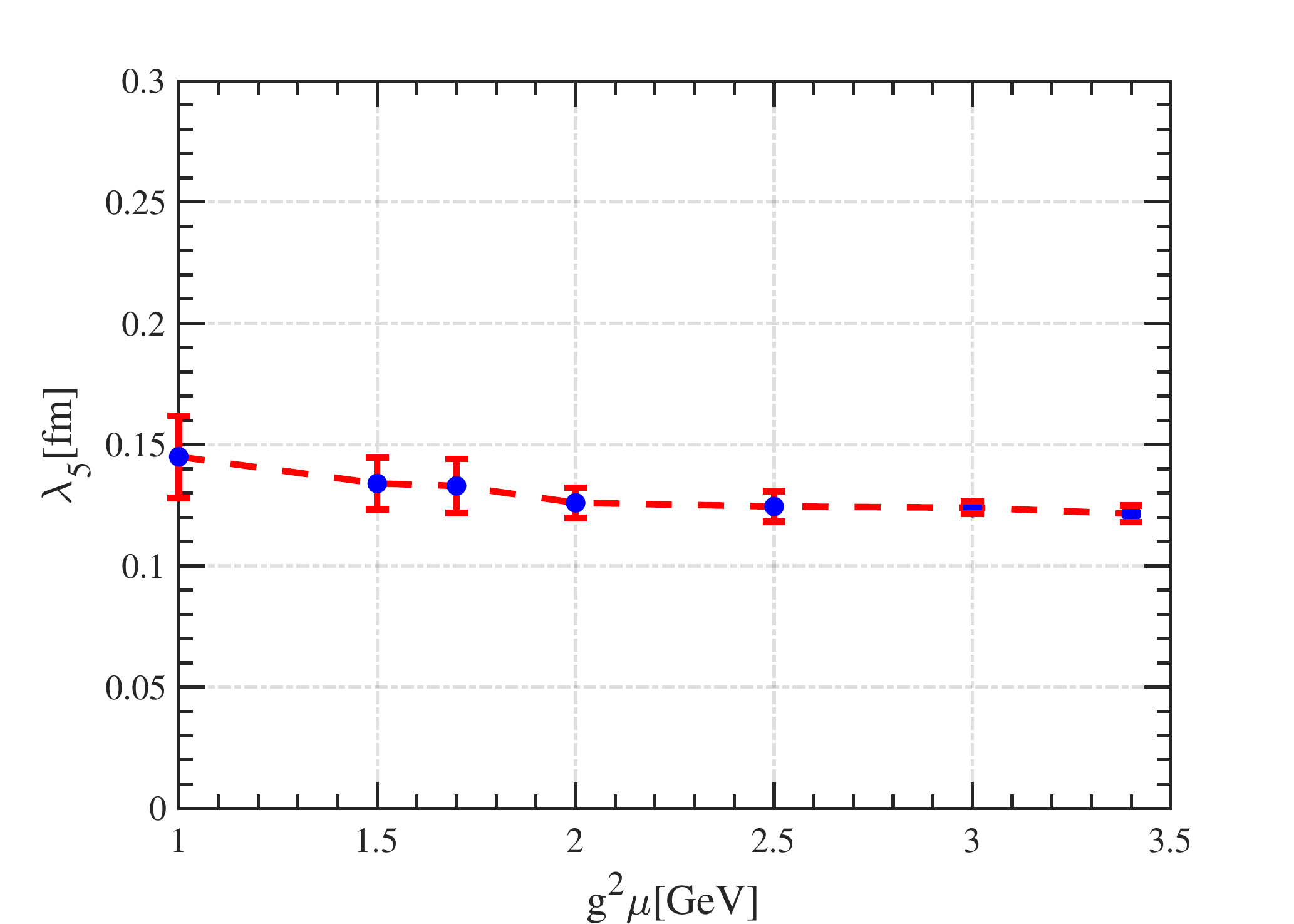}
		\caption{{\it Upper panel.} Dimensionless correlation length of the chiral charge density $n_5$ versus $g^2\mu$. 
		{\it Lower panel.} Correlation length of $n_5$ in physical units versus $g^2\mu$.
		Lattice parameters are the same as in Fig.~\ref{fig5}. }
		\label{fig8}
\end{figure}

From the results in Fig.~\ref{fig5} we can estimate the correlation length, $\lambda_5$, of $n_5$ by looking for the value of $|x_\perp|$
such that the correlator decays to $1/e$ of its value at $x_\perp=0$;
we follow the same procedure depicted in the previous subsection for the definition and computation of $\lambda_T$.
We summarize the results of this estimate in Fig.~\ref{fig8}
where we plot $\lambda_5$ versus $g^2\mu$,
both as a dimensionless quantity (upper panel) and in physical units (lower panel).
The correlation length sits in the range $(0.5,2.5)$ in units of $g^2\mu$, in agreement with the results we have found for 
the topological charge: the correlations of $\rho_T$ are transmitted
to those of $n_5$ almost unaffected.

As check of our results, we have studied how the correlation lengths change with the lattice spacing. For example, we have considered $g^2\mu=2$ GeV and changed the lattice spacing from $a=4/91$ fm to $a=4/181$ fm: this changes $\lambda_5$ from $g^2\mu\lambda_5=1.1$ to $g^2\mu\lambda_5=0.8$. Similarly, for  $g^2\mu=3.4$ GeV we have changed the lattice spacing from $a=4/91$ fm to $a=4/181$ fm: this changes $\lambda_5$ from $g^2\mu\lambda_5=2$ to $g^2\mu\lambda_5=1.74$. 
These checks show that our results are quite reliable.

\section{Conclusion and outlook}\label{sec5}
We have studied the correlations of the topological charge density, $\rho_T$, carried by the strong gluon fields
in the early stage of high energy nuclear collisions; besides, we have analyzed the production and the correlations
of the chiral density per unit rapidity in the transverse plane, $n_5$, 
produced by the chiral anomaly of QCD. In fact in the early stage of the collisions,
assuming the McLerran-Venugopalan initialization and the Glasma picture, 
$\rho_T\propto\bm E \cdot \bm B \neq 0$ and $dn_5/d\tau=\tau\rho_T$,
where $\bm E$ and $\bm B$ denote the color electric and color magnetic fields and $\tau$ is the proper time.
We have described the evolution of the gluon fields by means of the Classical Yang-Mills (CYM) equations,
that we have solved numerically for the case of SU(2):
we have followed the evolution up to $\tau=1$ fm/c, in agreement with the switch time from the
 CYM evolution to relativistic hydro or transport in nucleus-nucleus collisions
at the RHIC energy \cite{Ruggieri:2013bda,Ruggieri:2013ova,Ruggieri:2015tsa,Ruggieri:2015yea,Gale:2012rq,Gale:2012in}.
We have analyzed the case of a longitudinal boost invariant expansion that is more relevant
for the modeling of the early stage of the collisions. 
Despite some amount of information available in the literature on the formation of topological charge domains in the early
stage of high energy nuclear collisions, see for example~\cite{PhysRevD.98.014025,PhysRevD.97.034034,Lappi:2006fp},
a concrete calculation that takes into account the full evolution of the gluon field
with the initial condition given by the MV model
is still missing: we aim to cover this lack of information here.
Besides its theoretical own interest, this study is relevant for the phenomenology of high energy nuclear collisions:
for example, just recently it has been pointed out that the production of chiral density in the early stage can
affect the polarization of the $\Lambda$ particles \cite{PhysRevC.82.057902,PhysRevD.98.071902}
and puts constraints on the strength of the magnetic field produced in the collisions: it is therefore important to have both a qualitative and
a quantitative understanding of the $n_5$ that is produced in the early stage where classical gluon fields dominate the dynamics.

We have computed the correlator of $\rho_T$ in the transverse plane, studying how the correlation develop and 
how the correlation length depends on $g^2\mu$, namely on the density of color charges in the transverse plane:
this number is the only energy scale in the initial condition, and increasing it amounts to model a collision with a higher 
center of mass energy.
Examining the range $(1,3.4)$ GeV for $g^2\mu$, we have found, for the correlation length $\lambda_T$,
that $\lambda_T\approx O(1)\times 1/g^2\mu$: increasing the collision energy amounts to fit more correlated domains
within the transverse plane, the density of these being given approximately by $O(1)\times (g^2\mu)^2$.
These domains remain microscopic with respect to the nucleon, in the sense that in physical units the correlation length
sits in between the $10\%$ and the $20\%$ of the proton radius.
The correlators and correlation lengths that we have found agree with previous studies \cite{PhysRevD.97.034034,PhysRevD.98.014025,PhysRevD.88.031503,PhysRevD.97.076004,DUMITRU20147}. 

We have also studied the production of the chiral density in the transverse plane per unit of rapidity, 
$n_5\equiv dN_5/d\eta d^2x_\perp$, that is formed by the chiral anomaly of QCD. For this quantity,
we have analyzed the production rate and how it depends on $g^2\mu$, giving some estimate
of this value when the system reaches a steady state. 
The $\langle n_5^2\rangle$ depends on $g^2\mu$, which is obvious since increasing the latter amounts to increase
the magnitude of the gluon fields in the initial condition. We have estimated the proper time to reach a steady state, $\tau_s$,
in the production of $n_5$ to be $\tau_s\approx 1/g^2\mu$. To give concrete numbers, 
we have found
$\sqrt{\langle N_5^2\rangle/V^2}\approx 1.79\times 10^{-1}~\rm fm^{-3}$ at $g^2\mu= 1.0~\rm GeV$, 
$\sqrt{\langle N_5^2\rangle/V^2}\approx 8.85\times 10^{-1}~\rm fm^{-3}$ at $g^2\mu= 1.7~\rm GeV$,
$\sqrt{\langle N_5^2\rangle/V^2}\approx 6.48~\rm fm^{-3}$ at $g^2\mu= 3.4~\rm GeV$.
The average chiral density per unit rapidity can be quite large 
for the largest value of $g^2\mu$ that we have used, at least until the steady state is formed:
the longitudinal expansion will dilute $\langle N_5^2\rangle/V^2$ as $1/\tau^2$ in the steady state, 
when the topological charge density is low enough that no substantial new $n_5$ is produced.

We have analyzed the correlation domains in the transverse plane and how their size is affected by $g^2\mu$:
as for $\rho_T$ we have found that the correlation length of $n_5$, namely $\lambda_5\approx O(1)\times 1/g^2\mu$,
and within numerical uncertainty $\lambda_5 \approx \lambda_T$.

The qualitative picture that arises from our study is that increasing the collision energy, the transverse plane
gets populated by a larger amount of correlated domains of $\rho_T$ and $n_5$, the size of these becoming smaller as
$\simeq 1/g^2\mu$ and their density increasing as $(g^2\mu)^2$; chiral density per unit rapidity forms immediately after the collision,
and a steady state is achieved after a short proper time range $\tau_s\approx (0.1~\mathrm{fm/c},0.2~\mathrm{fm/c})$,
the lowest value corresponding to $g^2\mu=3.4$ GeV and the highest to $g^2\mu=1$ GeV.
We mention that we have also also found some anticorrelation at the initial time, in agreement with
previous studies of the gauge invariant correlators of the gluon fields
\cite{PhysRevD.88.031503,PhysRevD.97.076004}, but the evolution cancels this anticorrelation 
in a time range $g^2\mu \tau=O(1)$.

A natural extension of the work reported here is the study of the correlations in rapidity, introducing fluctuations 
along the longitudinal direction in the initial condition; besides, the systematic study we have presented here
can be enriched by modeling the realistic transverse plane geometries of nucleus-nucleus
and proton-nucleus collisions. 
 In addition to this, the diffusion of the chiral density
in the transverse plane in the early stage might be an interesting topic to investigate. 
Even more, the production of photons in the early stage due to 
coexistence of $n_5$ and an electromagnetic field  in the early stage is worth 
of being studied. We plan to address
these topics in the near future.

\begin{acknowledgments}	 
The authors acknowledge Navid Abbasi, Marco Frasca and John Petrucci for inspiration, 
discussions and comments on the first version of this article.
M. Ruggieri is supported by the National Science Foundation of China (Grants No.11805087 and No.11875153) and by the Fundamental Research Funds for the Central Universities (Grant number 862946).
The work of J. H. Liu is supported by China Scholarship Council (scholarship number 201806180032).  
H. F. Zhang is supported by the National Science Foundation of China (Grant No.11675066).
\end{acknowledgments}

\bibliography{Topological_charge}

\end{document}